\documentclass[letterpaper, 10 pt, conference]{ieeeconf} 
\IEEEoverridecommandlockouts
\usepackage{amsmath,amssymb,amsfonts}
\usepackage{algorithmic,algorithm}
\usepackage{graphicx}
\usepackage{subcaption}
\usepackage{xcolor}
\usepackage{etoolbox}
\usepackage{mathtools}
\usepackage{stix}
\usepackage{epstopdf}
\usepackage{xparse}
\usepackage{subcaption} 
\def\BibTeX{{\rm B\kern-.05em{\sc i\kern-.025em b}\kern-.08em
    T\kern-.1667em\lower.7ex\hbox{E}\kern-.125emX}}
\usepackage{tikz}
\usetikzlibrary{arrows.meta,calc,fit,shapes.geometric}

\graphicspath{ {images/} }

\newtheorem{assume}{Assumption}
\newcommand{\R}{\mathbb{R}}

\newtoggle{andrew}
\newtoggle{jack}
\togglefalse{andrew}   
\togglefalse{jack}

\newcommand{\andrew}[1]{\iftoggle{andrew}{\textcolor{blue}{#1}}{#1}}
\newcommand{\jack}[1]{\iftoggle{jack}{\textcolor{blue}{#1}}{#1}}

\begin{document}
\title{Fuel-Efficient Formation Flying via Constrained Extremum Seeking}

\author{Kwang Hak Kim$^{\dag}$, Luis G. Crespo$^{\star}$ and Michael J. Acheson$^{\star}$
\thanks{This work was supported by the Doolittle project of the US Navy.}
\thanks{$^{\dag}$Department of Mechanical and Aerospace Engineering, UC San Diego, 9500 Gilman Drive, La Jolla, CA, 92093, USA}
\thanks{$^{\dag}$Corresponding author, {\tt\small kwk001@ucsd.edu}.}
\thanks{$^{\star}$Vehicle Dynamics \& Controls Branch, NASA Langley Research Center, Hampton, VA, 23681, USA.}
}

\maketitle

\begin{abstract}  
This paper presents a model-free autonomous guidance architecture for minimal drag of a trailing aircraft in formation flight. Realizing the fuel savings resulting from such a formation requires learning and tracking a time-varying vortex while avoiding locations where the aircraft experiences an abrupt rolling moment disturbance. A guidance architecture for fuel-efficient safe flight, called Adaptive Plateau-seeking Optimal Guidance with ModulatEd Excitations (APOGEE), is proposed herein. Key features of APOGEE are its ability to modulate the amplitude of the excitation required to learn the unknown objective and constraint functions, to adaptively adjust the learning rates, and to keep the flight safe during the search for the optimal formation. A threshold-gated modulation reduces the amplitude of excitation when the objective function reaches an acceptably large value, thus improving flying qualities and further reducing fuel consumption. Furthermore, a Newton-based learning rule, whose estimate of the inverse of the Hessian is spectrally clamped, adapts the learning rates to the local curvature of the objective function in order to accelerate convergence and lower the chance of overshooting the optimum. Finally, a safety filter having an unknown constraint function is used to prevent a loss of control event prompted by the above mentioned disturbance. The efficacy of APOGEE is demonstrated by using a high-fidelity model of an F-16 jet and an experimentally obtained vortex model. Being model-free, APOGEE does not require modeling the dependence of the unknown functions on the characteristics of the leading aircraft, the flight condition, the crosswind, and the vortex-induced turbulence. 
\end{abstract}

\section{Introduction}
\begin{figure}[t!]
\centering
\includegraphics[trim={0cm 0cm 0cm 0cm},clip,width=.45\textwidth]{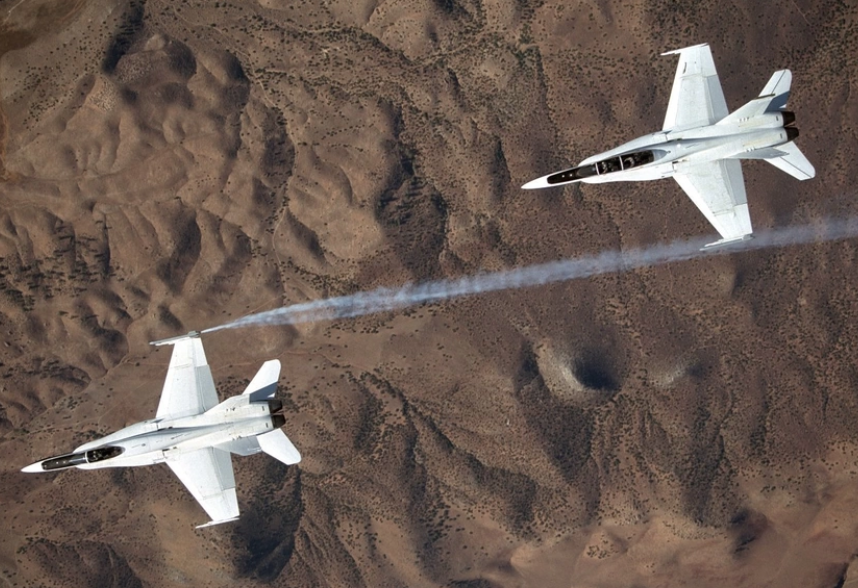}
\caption{Formation flying configuration, NASA Dryden Flight Research Center.}
\label{ff}
\vspace{-0.6cm}
\end{figure}

Formation flying, as illustrated in Fig.~\ref{ff}, can reduce the fuel consumption of a follower aircraft by ``riding'' the vortex created by a leading aircraft. Flight experiments performed at NASA Armstrong showed that the vortex-induced rolling moment is a good proxy for minimum drag~\cite{Hansen02}. This moment is estimated using the equations of motion and the measured states after subtracting the rolling moment corresponding to free flow for such a flight condition. These flight experiments also showed that flying at an optimal relative position reduces drag \andrew{by up to} 22\%, with corresponding reductions in fuel consumption measured in flight~\cite{vachon2002f}, and extensions to transport-class aircraft in~\cite{pahle2012initial}. The benefits of wake-riding degrade gradually with increased nose-to-tail distances after peaking at three span lengths aft\andrew{~\cite{Hansen02}}. 

Realizing these savings autonomously requires tracking an unknown and potentially time-varying optimal position whose location depends on the mass, state and geometry of the leading aircraft, the relative wind and the orientation of the trailing aircraft. Wake-model-based designs estimate the vortex state from onboard measurements, using a Kalman filter built on a vortex model of the leader's wake~\cite{hemati2014wake,brodecki2015autonomous} or observers derived from similar wake parameterizations~\cite{devries2016wake}. However, these estimates are only as good as the assumed model. Even in a simplified setting, the wake nonlinearity biases Kalman-type estimates and the filter diverges in static formation~\cite{hemati2014wake}, thereby motivating the development of model-free methods that only use instantaneous measurements of the unknown functions.

One model-free framework is peak-seeking control~\cite{chichka2006peak,ryan2010peak}, in which the local gradient and Hessian of the measured performance map are estimated by a Kalman filter over commanded excitation maneuvers and the position is stepped toward the estimated peak. Peak-seeking has been extensively developed through flight tests, including trim optimization for reduced fuel consumption~\cite{schaefer2013peak,brown2013peak}. Its estimates, however, rest on a local quadratic parameterization of the map and on noise covariances that must be tuned to an unknown environment, and the approach does not carry practical closed-loop stability guarantees. 

In contrast, Extremum Seeking (ES) extracts the gradient of an unknown function by demodulating a periodic dither, a structure for which stability has been established by averaging and singular perturbation~\cite{Krstic_firstESproof_2000}. Dither-based ES has been developed into an extensive model-free optimization framework~\cite{ariyur_krstic_ESCbook_2003}, with a century of developments surveyed in~\cite{scheinker_100yrs}. Three extensions are of particular relevance. 

\jack{\andrew{The first extension of note is}} the Newton-based ES, introduced for scalar maps in~\cite{moase2010newton}, generalized through derivative estimation in~\cite{nesic2010unifying}, and extended to the multi-variable case in \cite{ghaffari_newton_MV_2012,todorovski2024newton}. This method removes the dependence of the convergence rate on the unknown Hessian of the map. The second extension is Practically Safe ES, which combines ES with Control Barrier Function (CBF) safety filters~\cite{ames2016_CBFQP, kim2025robust} to prevent the violation of an unknown constraint through quadratic program (QP)-style filters~\cite{alan_semiglob_safeESC_2025,alan_local_assign_safeESC} or logarithmic barriers \cite{qixu_logbarrier_ESC_2026}. Finally, existing excitation-modulating ES designs~\cite{yilmaz2025exponential,suttner2019extremum} generally remove the residual oscillation through unbounded demodulation gains, probing frequencies, or update rates. However, a noisy objective measurement and rate-limited signals in the inner loop pose significant challenges to such approaches.

ES has been used in a wide range of applications thanks to its robustness, efficacy, and minimal modeling requirements~\cite{ghaffari2014power, killingsworth2009hcci}. Classical gradient-based ES has been applied in~\cite{Binetti_FormationFlight_ESC_2003} for the Lockheed C-5 aircraft with the follower's pitch angle as the performance output. \andrew{The design in~\cite{brodecki2015autonomous} pairs ES with an extended Kalman filter, introducing a wake model with uncertain parameters}, and~\cite{ryan2018formation} uses an objective function that combines several performance signals at the cost of potentially introducing delayed and uncertain measurements. Nonetheless, none of these designs bound the transient response leading to the safety-critical roll reversal, nor do they eliminate the dither once an acceptable objective value is reached, a critical aspect of any fuel-efficient strategy~\cite{chichka2006peak}.

This paper presents the Adaptive Plateau-seeking Optimal Guidance with ModulatEd Excitations (APOGEE) algorithm, an autonomous guidance law for the follower aircraft. APOGEE uses a Newton-based learning rule that speeds up convergence, a practical CBF-QP safety filter that recovers the aircraft from incursions past a safety buffer before the \andrew{downwash region} is reached, and a threshold-gated amplitude modulation that saves fuel by switching the excitation on and off based on the value of the objective function. These modifications address practical problems that the \jack{aforementioned works} do not cover, such as the unbounded transient on a strongly asymmetric function or the fuel cost of the dithering maneuver itself. The guarantees behind each mechanism are local and practical, and no formal result covers their composition. APOGEE inherits the underlying assumptions of ES theory: a measurable proxy for the true objective, a potentially \andrew{non-stationary, drifting vortex}, an identifiable inner-loop lag at the excitation frequencies, and sufficient separation between the excitation, estimation, and update time scales.

\section{Guidance Architecture and Optimization Configuration}

\jack{\andrew{The unknown map defining the objective or constraint has a physical interpretation; for example, maximizing the selected objective should approximately minimize drag.\footnote{Drag is difficult to measure both accurately and promptly, thus requiring a proxy signal varying at a time scale comparable to the rotational aircraft dynamics.}} The control engineer must choose the independent variables of this function; these variables must meet a few requirements. First, holding the variables constant must settle the closed loop to a stable equilibrium so that a static map from input to the measured output exists in steady state.} A fixed relative position, for example, satisfies this condition, but an arbitrarily fixed pair of nonzero pitch and roll angles does not. Second, the inner loop \jack{must quickly track the commanded reference at a higher rate than the probing signal, such} that the plant reaches a quasi-steady state on the time scale of the perturbation. Third, the steady-state map from the chosen inputs to the measured output must admit a single extremum of known type, with nonzero curvature at that point.

We consider a two-aircraft formation in the North-East-Down (NED) convention, with the trailing aircraft's position parameterized by the lateral and vertical offsets relative to \andrew{an arbitrary reference position}
\begin{align}
    p = \begin{bmatrix}
        y- \andrew{y_{\mathrm{ref}}} \\
        z -\andrew{z_{\mathrm{ref}}}
    \end{bmatrix} \in \R^2\,.
\end{align}
Obtaining precise measurements of $y$ and $z$ relative to the leading aircraft or ground references is notoriously difficult. However, APOGEE is reference-agnostic, eliminating the need for advanced sensors or complex online measurement methods. The use of the notation ``$y$'' and ``$z$'' is strictly for conceptual clarity and does not imply a hardware dependence on precise relative measurement. Consequently, these \andrew{reference positions} can be set to zero upon starting APOGEE. 

\subsection{Guidance and Control Architecture}

\begin{figure}[t!]
\centering
\begin{tikzpicture}[
  x=1cm, y=1cm,
  font=\scriptsize,
  line width=0.5pt,
  blk/.style={draw, rectangle, rounded corners=1pt, align=center,
              minimum height=0.8cm, inner sep=2.5pt},
  sig/.style={-{Latex[length=1.5mm]}},
  dot/.style={fill, circle, inner sep=0pt, minimum size=2.6pt}
]
\draw[dashed, line width=1pt, rounded corners=3pt, draw=blue!80] (3.65,0.55) rectangle (8.2,2.5);
\node[anchor=north, font=\small, text=blue!80] at (6,2.5) {Inner closed-loop};
\node[blk, minimum width=0.8cm] (pd)   at (1.0,1.6) {PD};
\node[blk, minimum width=0.8cm] (sat)  at (2.9,1.6) {};
\draw ([xshift=-0.28cm,yshift=-0.18cm]sat.center) -- ([xshift=-0.10cm,yshift=-0.18cm]sat.center)
   -- ([xshift= 0.10cm,yshift= 0.18cm]sat.center) -- ([xshift= 0.28cm,yshift= 0.18cm]sat.center);
\node[blk, minimum width=1.9cm] (ctrl) at (4.9,1.6) {LQG $+$ $\mathcal{L}_1$\\ adaptive controller};
\node[blk, minimum width=1.5cm] (f16)  at (7.3,1.6) {F-16 $+$\\ vortex model};
\draw[sig] (pd)   -- node[above]{$\phi_{\mathrm{cmd}},\theta_{\mathrm{cmd}}$} (sat);
\draw[sig] (sat)  -- (ctrl);
\draw[sig] (ctrl) -- node[above]{$u$} (f16);
\draw (f16.south) -- (7.3,0.8) -- node[above, inner sep=1pt]{system states} (4.9,0.8);
\node[dot] at (4.9,0.8) {};
\draw[sig] (4.9,0.8) -- (ctrl.south);
\draw (4.9,0.8) -- (1.0,0.8);
\draw[sig] (1.0,0.8) -- (pd.south);
\node[blk, minimum width=1.4cm] (apogee) at (4.4,-0.1) {APOGEE};
\draw (f16.east) -- (8.5,1.6) -- (8.5,-0.1);
\draw[sig] (8.5,-0.1) -- node[above,inner sep=1pt]{measured output} (apogee.east);
\draw (apogee.west) -- node[above,pos=0.12]{$p_\mathrm{cmd}$} (0.1,-0.1) -- (0.1,1.6);
\draw[sig] (0.1,1.6) -- (pd.west);
\end{tikzpicture}
\caption{\jack{Closed-loop system with APOGEE as outer-loop guidance.}}
\label{fig:inner-loop}
\vspace{-0.5cm}
\end{figure}
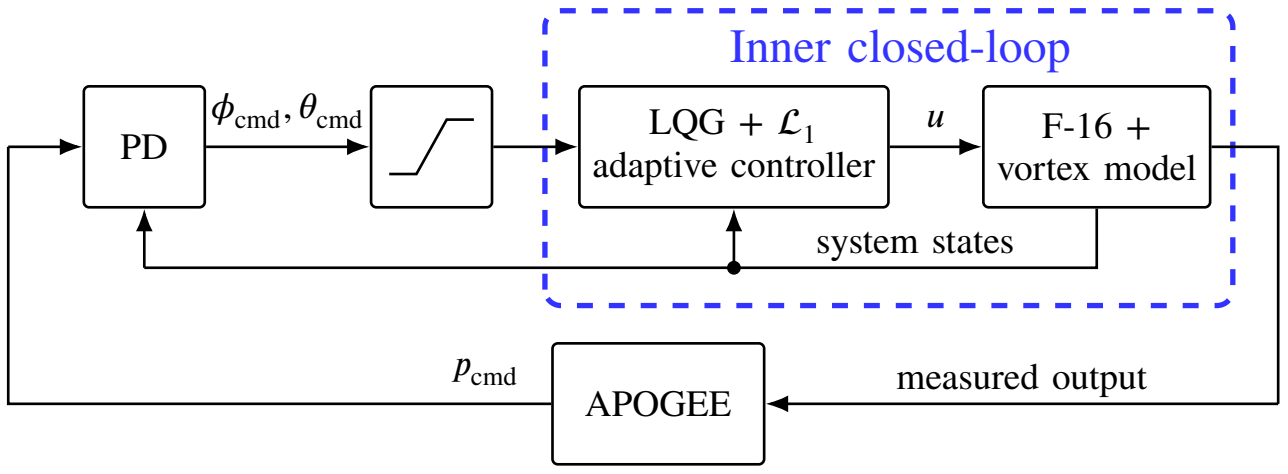

\jack{APOGEE operates as an outer loop that generates a position command $p_\mathrm{cmd}$ as the relative cross-track and vertical reference errors, as shown in Fig.~\ref{fig:inner-loop}. A PD controller maps $p_\mathrm{cmd}$ to roll and pitch commands ($\phi_{\mathrm{cmd}}$, $\theta_{\mathrm{cmd}}$), which are range limited before entering the inner closed loop. There, an LQG controller with $\mathcal{L}_1$ adaptive augmentation generates the aircraft control inputs. The design of the inner loop is not the subject of this paper, and it is treated as a black box hereafter.}

ES \andrew{estimates} the gradient and the Hessian of the unknown map by commanding a sinusoidal dither with frequencies $\omega_y$ and $\omega_z$ in $y$ and $z$ respectively.  The \jack{dither} passes through the reference-generation and tracking cascade before reaching the vehicle, which responds with an additional phase lag. Thus, the closed-loop response is delayed relative to the command. Without compensation, this delay can lead to erroneous estimates. Hence, we identify the estimated delay offline and shift the demodulation phase by
\begin{align}\label{eq:phase_lag}
    \beta_i \coloneqq -\omega_i \tau_i\,, 
\end{align}
where $i$ is either $y$ or $z$, and $\tau_i$ is the inner-loop delay for the corresponding dither frequency. So the demodulation signals are aligned with the commanded dither.

\subsection{Unconstrained Optimization}
While our objective is to reduce fuel consumption, drag is not directly measurable and difficult to estimate~\cite{ryan2018formation}. Instead, we adopt the observation made in~\cite{Hansen02}, which approximates the minimum drag configuration to the location where the vortex-induced rolling moment coefficient $C_l$ is maximal. Our goal is to solve
\begin{align}\label{eq:optimization_problem}
    p^* = \arg \max_{p} \; C_l(p)\,,
\end{align}
only using real-time measurements. To this end we state the following assumption.
\begin{assume}\label{assume:map}
The map $C_l : \R^2 \to \R$ is twice continuously differentiable, bounded, and admits a local maximizer $p^* \in  \R^2$, satisfying $\nabla C_l(p^*) = 0$.
\end{assume}
It is important to note that there is a small offset between the maximizer for $C_l$ and the minimizer of drag. Objective functions built from a single metric approximate the drag-reduction map acceptably well, \jack{whereas those using a weighted sum of several metrics~\cite{ryan2018formation}} might sharpen that approximation. However, the potential delay and estimation error of these metrics might render them impractical.

Additionally, the vortex-induced rolling moment must be estimated in practice, for instance, by comparing the aircraft's response inside and outside the vortex using a validated aerodynamic model. Forming that residual by direct subtraction amplifies noise and ignores the additional degrees of freedom, whereas a Kalman-filter-based disturbance observer that fuses several sensor outputs with the coupled dynamics returns a smoother estimate. The real-time implementation of APOGEE uses a Kalman filter to approximate the map. Even though this practice imposes a modeling requirement, ES itself does not impose any limitation on the structure of the mapping.


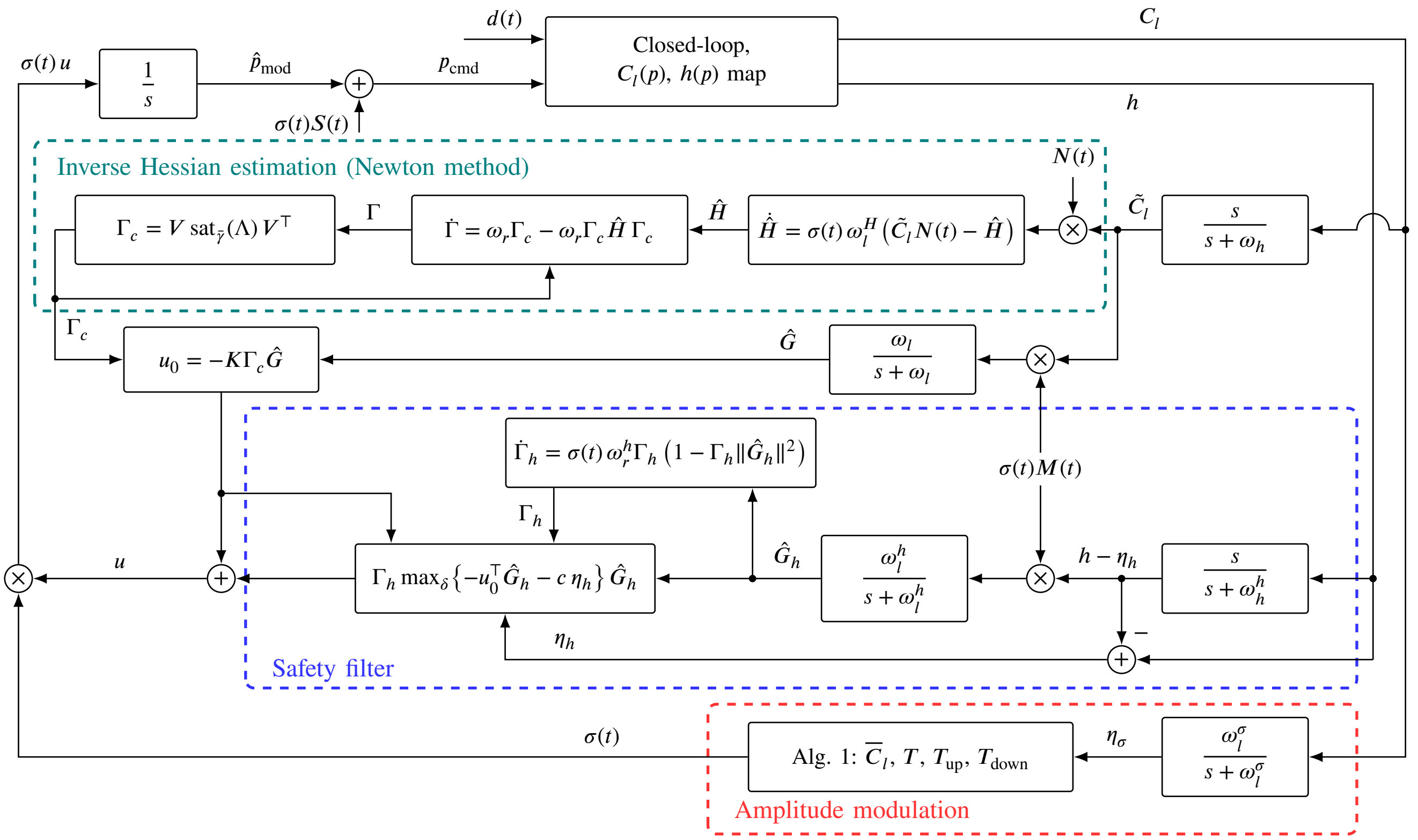
\begin{figure*}[t!]
\centering
\begin{tikzpicture}[
  x=1cm, y=1cm,
  font=\footnotesize,
  line width=0.5pt,
  blk/.style={draw, rectangle, rounded corners=1pt, align=center,
              minimum height=0.85cm, inner sep=3pt},
  sum/.style={draw, circle, inner sep=0pt, minimum size=0.34cm},
  mul/.style={draw, circle, inner sep=0pt, minimum size=0.34cm},
  sig/.style={-{Latex[length=1.7mm]}},
  dot/.style={fill, circle, inner sep=0pt, minimum size=2.6pt}
]

\node[blk, minimum width=1.2cm] (int) at (2.0,9.2) {$\dfrac{1}{s}$};
\node[sum] (s1) at (4.6,9.2) {$+$};
\node[blk, minimum width=3.6cm, minimum height=1.1cm] (plant) at (8.7,9.475)
      {Closed-loop,\\[1pt] $C_l(p)$, $h(p)$ map};
\draw[sig] (int) -- node[above]{$\hat p_{\mathrm{mod}}$} (s1);
\draw[sig] (s1) -- node[above]{$p_\mathrm{cmd}$} (s1 -| plant.west);
\draw[sig] ([xshift=-1cm,yshift=0.275cm]plant.west) -- node[above]{$d(t)$}
           ([yshift=0.275cm]plant.west);
\draw[sig] (4.6,8.6) -- (s1.south);
\node[left] at (4.6,8.7) {$\sigma(t) S(t)$};

\node[blk, minimum width=1.8cm] (wash) at (15.4,7.4) {$\dfrac{s}{s+\omega_h}$};
\draw[dashed, line width = 1pt, rounded corners=3pt, draw=teal] (0.6,6.4) rectangle (13.8,8.5);
\node[anchor=north west, font=\small, text=teal] at (0.75,8.42) {Inverse Hessian estimation (Newton method)};
\node[mul] (mN) at (13.4,7.4) {$\times$};
\node[blk] (lpH) at (11.1,7.4)
      {$\dot{\hat H}=\sigma(t)\,\omega_l^{H}\bigl(\tilde{C}_lN(t)-\hat H\bigr)$};
\node[blk, minimum width=3.4cm] (ricc) at (6.95,7.4)
      {$\dot\Gamma=\omega_r\Gamma_c-\omega_r\Gamma_c\hat H\,\Gamma_c$};
\node[blk, minimum width=3.2cm] (clamp) at (2.7,7.4)
      {$\Gamma_c = V\,\mathrm{sat}_{\bar\gamma}(\Lambda)\,V^\top$};
\draw[sig] (wash.west) -- node[above,pos=0.3]{$\tilde C_l$} (mN.east);
\node[dot] at (13.95,7.4) {};
\draw[sig] (13.4,8.05) -- (mN.north);
\node[above] at (13.4,8.05) {$N(t)$};
\draw[sig] (mN.west) -- (lpH.east);
\draw[sig] (lpH.west) -- node[above]{$\hat H$} (ricc.east);
\draw[sig] (ricc.west) -- node[above]{$\Gamma$} (clamp.east);

\draw (clamp.west) -- (0.85,7.4);
\draw[sig] (0.85,7.4) -- (0.85,6.55) -- (6.95,6.55) -- (ricc.south);
\node[dot] at (0.85,6.55) {};

\node[mul] (mM) at (13.0,5.8) {$\times$};
\node[blk, minimum width=1.8cm] (lpG) at (11.3,5.8) {$\dfrac{\omega_l}{s+\omega_l}$};
\node[blk, minimum width=2.4cm, minimum height=0.8cm] (u0) at (2.9,5.8)
      {$u_0 = -K\Gamma_c\hat G$};
\draw[sig] (13.95,7.4) -- (13.95,5.8) -- (mM.east);
\draw[sig] (mM.west) -- (lpG.east);
\draw[sig] (lpG.west) -- node[above,pos=0.08]{$\hat G$} (u0.east);

\draw[sig] (0.85,6.55) -- (0.85,5.8) -- (u0.west);
\node[right] at (0.87,6.17) {$\Gamma_c$};

\draw[dashed, line width = 1pt, rounded corners=3pt, draw=blue!80] (3.2,1.75) rectangle (16.9,5.2);
\node[anchor=south west, font=\small, text=blue!80] at (3.4,1.7) {Safety filter};

\node[blk, inner xsep=6pt] (sfy) at (6.4,3.1)
      {$\Gamma_h\max_\delta\!\left\{-u_0^{\top}\hat G_h - c\,\eta_h\right\}\hat G_h$};
\node[blk, anchor=west] (gh) at (6.4,4.65)
      {$\dot\Gamma_h=\sigma(t)\,\omega_r^{h}\Gamma_h\left(1-\Gamma_h\|\hat G_h\|^2\right)$};
\node[blk, minimum width=1.8cm] (lph) at (11.2,3.1) {$\dfrac{\omega_l^h}{s+\omega_l^h}$};
\node[mul] (mh) at (13.0,3.1) {$\times$};
\node[blk, minimum width=1.8cm] (hpf) at (15.4,3.1) {$\dfrac{s}{s+\omega_h^h}$};
\node[sum] (s3) at (14.0,2.1) {$+$};

\node[inner sep=2pt] (sM) at (13.0,4.45) {$\sigma(t)M(t)$};
\draw[sig] (sM.north) -- (mM.south);
\draw[sig] (sM.south) -- (mh.north);

\draw[sig] (hpf.west) -- node[above]{$h-\eta_h$} (mh.east);
\draw[sig] (mh.west) -- (lph.east);
\draw[sig] (lph.west) -- node[above,pos=0.2]{$\hat G_h$} (sfy.east);
\node[dot] at (9.45,3.1) {};
\draw[sig] (9.45,3.1) -- (9.45,0 |- gh.south);

\draw[sig] (7.0,0 |- gh.south) -- node[left]{$\Gamma_h$} (7.0,0 |- sfy.north);

\node[dot] at (14.0,3.1) {};
\draw[sig] (14.0,3.1) -- (s3.north);
\node[right] at (14.02,2.4) {$-$};
\draw[sig] (s3.west) -- node[above,pos=0.9]{$\eta_h$} (6.4,2.1) -- (sfy.south);

\node[sum] (s2) at (2.9,3.1) {$+$};
\draw[sig] (u0.south) -- (s2.north);
\node[dot] at (2.9,4.15) {};
\coordinate (u0in) at ([xshift=0.45cm]sfy.north west);
\draw[sig] (2.9,4.15) -- (2.9,4.15 -| u0in) -- (u0in);
\draw[sig] (sfy.west) -- (s2.east);

\node[mul] (msig) at (0.4,3.1) {$\times$};
\draw[sig] (s2.west) -- node[above]{$u$} (msig.east);
\draw[sig] (msig.north) -- (0.4,9.2) -- (int.west);
\node[right] at (0.3,9.45) {$\sigma(t)\, u$};

\draw[dashed, line width = 1pt, rounded corners=3pt, draw = red!80] (8.9,-0.05) rectangle (16.9,1.55);
\node[anchor=south west, font=\small, text = red!80] at (9.1,-0.05) {Amplitude modulation};
\node[blk, minimum width=1.8cm] (modlp) at (15.4,0.9) {$\dfrac{\omega_l^\sigma}{s+\omega_l^\sigma}$};
\node[blk, minimum width=4.0cm] (alg) at (11.4,0.9)
      {Alg.~\ref{alg:amp_mod}: $\overline{C}_l$, $T$, $T_{\mathrm{up}}$, $T_{\mathrm{down}}$};
\draw[sig] (modlp.west) -- node[above]{$\eta_\sigma$} (alg.east);
\draw[sig] (alg.west) -- node[above,pos=0.2]{$\sigma(t)$} (0.4,0.9) -- (msig.south);

\draw ([yshift=0.275cm]plant.east) -- node[above,pos=0.55]{$C_l$} (17.5,9.75) -- (17.5,7.4);
\node[dot] at (17.5,7.4) {};
\draw[sig] (17.5,7.4) -- (17.5,0.9) -- (modlp.east);
\draw[sig] (17.5,7.4) -- (17.3,7.4) arc[start angle=0, end angle=180, radius=0.2] -- (wash.east);

\draw (plant.east |- s1) -- node[below,pos=0.55]{$h$} (17.1,9.2) -- (17.1,2.1);
\draw[sig] (17.1,2.1) -- (s3.east);
\node[dot] at (17.1,3.1) {};
\draw[sig] (17.1,3.1) -- (hpf.east);

\end{tikzpicture}
\caption{Block diagram architecture of APOGEE. $d(t)$ represents unknown disturbances and measurement noise. A scaling of $1000$ applied to the output $C_l$ is omitted here for clarity (see Section~\ref{sec:sim}).}
\label{fig:block_diagram}
\vspace{-0.3cm}
\end{figure*}

\section{Extremum Seeking Design}\label{sec:ESC}
The key components of APOGEE, whose block diagram is depicted in Fig.~\ref{fig:block_diagram}, are introduced next.

\subsection{Classical ES}\label{sec:classical_ES}
Let $\hat{p}$ denote an estimate of $p^*$, with $p^*$ possibly time-varying. The commanded reference is the estimate plus a two-dimensional dither,
\begin{align}\label{eq:ESC_reference}
    p_\mathrm{cmd}(t) = \hat{p}(t) + S(t)\,,
\end{align}
where
\begin{align}\label{eq:S}
    S(t) = \begin{bmatrix}
        a_y \sin(\omega_y t)\\
        a_z \sin(\omega_z t)
    \end{bmatrix},
\end{align}
with $a_i >0$,  and $\omega_y \neq \omega_z$. The objective $C_l(p)$ is first high-pass filtered to remove its DC component, by using
\begin{align}\label{eq:HP_classic}
    \dot{\eta} = \omega_h \tilde{C}_l\,,
\end{align}
where $\tilde{C}_l = C_l - \eta$ and $\omega_h > 0$. Note that $\eta$ is a slow-varying estimate of $C_l$ at the current value of $p$. This signal is then demodulated and low-pass filtered using
\begin{align}\label{eq:LP_classic}
    \dot{\hat G} &= \omega_l\left(\tilde{C}_l M(t) - \hat G\right),
\end{align}
where $\omega_l >0$, and 
\begin{align}\label{eq:M}
    M(t) = \begin{bmatrix}
        \dfrac{2}{a_y}\sin(\omega_y t + \beta_y)\\[0.35cm]
        \dfrac{2}{a_z}\sin(\omega_z t + \beta_z)
    \end{bmatrix}\,,
\end{align}
is the demodulation signal. The DC component of (\ref{eq:LP_classic}), $\hat G$, is an estimate of the gradient of $C_l$ evaluated at $p$.

\jack{The classical update law},
\begin{align}\label{eq:classical_update}
    \dot{\hat p} = K\hat G\,,
\end{align}
where $K = \mathrm{diag}(k_y\,,k_z) \succ 0$ is the update gain, is unsuitable for~\eqref{eq:optimization_problem}. Flight experiments and CFD analyses have shown that the average value of $C_l(p)$ over a 30s time window is strongly asymmetric~\cite{Hansen02}, with the norm of the gradient on one side of the peak being substantially greater than on the other side. Since the step magnitude scales proportionally with the gradient, a constant $K$ tuned for acceptable progress on the shallow side can overshoot significantly on the steep side, while a gain sized for the steep side converges impractically slowly on the shallow side. This feature motivates the use of the Newton-based ES algorithm~\cite{ghaffari_newton_MV_2012}, to be presented next.

\subsection{Newton-based ES}\label{sec:newton}
The estimation error of $p^*$ averaged over the time period of the dither\footnote{The least common multiple of the period of all dithered signals is used.}~\cite{Krstic_firstESproof_2000,ghaffari_newton_MV_2012} is defined as $e_{\mathrm{avg}} \coloneqq \hat{p}_{\mathrm{avg}} - p^*$. This error is driven by
\begin{align}\label{eq:classical_avg}
    \andrew{\dot{e}_{\mathrm{avg}} \approx K H e_{\mathrm{avg}}}\,,
\end{align}
where $H$ is the Hessian of the $C_l(p)$ map evaluated at $\hat{p}_{\mathrm{avg}}$. Hence, the convergence rate is prescribed by the eigenvalues of $KH$. We use the Newton scheme of~\cite{ghaffari_newton_MV_2012} to eliminate the dependency of $\dot{e}_{\mathrm{avg}}$ on $H$. This practice allows ES to adjust the size of the perturbations of $\hat{p}$ leading to $p^*$. \andrew{Namely, small perturbations occur in regions where the norm of the gradient is large, whereas large perturbations occur where the norm of the gradient is small.} As such, this update rule not only prevents ES from overshooting the maximum in the former setting, but also accelerates convergence to the maximum in the latter setting.

The algorithm estimates the Hessian of the objective function from the dither in the response by second-order demodulation with the matrix
\begin{equation}\label{eq:N}
    N(t) = \begin{bmatrix}
        \dfrac{16}{a_y^{2}}\!\left(s_y^{2} - \tfrac{1}{2}\right) & \dfrac{4\,s_y s_z}{a_y a_z} \\[10pt]
        \dfrac{4\,s_y s_z}{a_y a_z} & \dfrac{16}{a_z^{2}}\!\left(s_z^{2} - \tfrac{1}{2}\right)
    \end{bmatrix}\,,
\end{equation}
where $s_i \coloneqq \sin(\omega_i t + \beta_i)$ with $i$ being $y$ or $z$. 
Hereafter we assume that the Hessian is negative definite and symmetric for all $p$ in the domain of interest.
As with the gradient estimate, the DC gain of the product $\tilde{C}_l N(t)$ carries Hessian information along with harmonic components. These harmonics are removed by a low-pass filter, thereby leading to
\begin{align}\label{eq:hess}
    \dot{\hat{H}} &= \omega_l^H\left(\tilde{C}_l N(t) - \hat H\right),
\end{align}
where $\omega_l^H > 0$. The integration of this equation yields the Hessian estimate $\hat{H}$.

The end goal of the Newton law is to formulate an update law leading to \andrew{$\dot{e}_{\mathrm{avg}} \approx -K e_{\mathrm{avg}}$} in steady state. The explicit inversion of the noisy $\hat{H}$ required to perform the desired cancellation is avoided by using the Riccati filter
\begin{equation}\label{eq:riccati_newton}
    \dot{\Gamma} = \omega_r \Gamma - \omega_r \Gamma \hat{H} \Gamma\,,
\end{equation}
for $\Gamma \in \R^{2\times 2}$ and $\omega_r > 0$.  This equation has an unstable equilibrium point at $\Gamma=0$ and a stable point where $\Gamma \prec 0$. When $\omega_r$ satisfies the time-scale separation condition in~\cite{ghaffari_newton_MV_2012} and $\Gamma(0)$ is both negative definite and symmetric, the integration of (\ref{eq:riccati_newton}) yields $\Gamma \to \hat H^{-1}$. This leads to the Newton update law
\begin{equation}\label{eq:newton_update}
    \dot{\hat{p}} = -K\,\Gamma \hat{G}\,,
\end{equation}
which reduces the averaged error dynamics to \andrew{$\dot{e}_{\mathrm{avg}} \approx -K e_{\mathrm{avg}}$} when $\Gamma \to H^{-1}$ is reached~\cite[Remark 2]{ghaffari_newton_MV_2012}.

\subsection{Spectral Clamping of the Inverse-Hessian Estimate}
The uniform rate sought by the Newton approach is only reached once $\Gamma$ converges to its asymptotic value. The transient rate, however, is governed by $-K\Gamma H$. As $\hat p$ crosses from the flatter to the steeper side of an asymmetric map, the gradient grows while $\Gamma$ still holds the flatter-side inverse curvature until the step overshoots $p^*$. Once the estimate is on the steeper side, $\hat H$ is relearned. Additionally, turbulence, measurement noise, and other uncertainties can drive the eigenvalues of $\hat H$ toward zero on a well-curved map. In either case, \eqref{eq:riccati_newton} can make a small estimated curvature yield an overly large transient gain. This outcome is prevented by bounding $\Gamma$.

\andrew{An eigenvalue decomposition of the Riccati state is $\Gamma = V\Lambda V^\top$, where $\Lambda =\mathrm{diag}\,(\lambda_y,\,\lambda_z)\in \mathbb{R}^{2 \times 2}$ is a diagonal matrix containing the eigenvalues of $\Gamma$, and $V \in \mathbb{R}^{2 \times 2}$ is an orthogonal matrix whose columns are the corresponding eigenvectors.} The desired bound is enforced by clamping the spectrum of $\Gamma$:
\begin{align}\label{eq:clamp}
    \Gamma_c \coloneqq V\;
    \begin{bmatrix}
        \mathrm{sat}(\lambda_y, -\bar{\gamma},0) & 0\\
        0 & \mathrm{sat}(\lambda_z, -\bar{\gamma},0)
    \end{bmatrix}\,
V^\top\,,
\end{align}
where $\bar{\gamma}>0$ and $\mathrm{sat}(\cdot,a,b)$ is the saturation function with lower limit $a$ and upper limit $b$.
The clamp is applied at the output of the Riccati equation and the clamped value is propagated both to the update law and back into the equation,
\begin{align}
    \dot{\Gamma} &= \omega_r \Gamma_c - \omega_r \Gamma_c \hat{H} \Gamma_c\,, 
    \label{eq:clampdyn}\\
    \dot{\hat{p}} &= -K\,\Gamma_c\,\hat{G}\,.\label{eq:clampdyn_update}
\end{align}
When an eigenvalue saturates, $\Gamma_c$ does not converge to $\hat{H}^{-1}$ and the averaged update law becomes \andrew{$-K \Gamma_c H$}. We therefore select $\bar\gamma$ to exceed the largest inverse curvature the map is expected to exhibit near $p^*$, so $\Gamma_c \rightarrow \Gamma$ near $p^*$. 
Hence, the spectrum of $\Lambda$ is only clamped during the transient response.
Near $p^*$ the clamp is inactive and \eqref{eq:clampdyn}--\eqref{eq:clampdyn_update} are equivalent to~\eqref{eq:riccati_newton}--\eqref{eq:newton_update}.

\subsection{Constrained Optimization}
The maximizer $p^*$ is close to a region where the rolling moment changes sign, \andrew{resulting in an abrupt roll moment disturbance and a loss of control}\footnote{Aircraft loss of control can be described as a response that is not predictably altered by commands, driven by nonlinear effects and coupling, and characterized by disproportionately large responses to small changes in the vehicle's state. Such a response compromises the ability to maintain heading, altitude, and wings-level flight.}. This outcome might be triggered by a moving vortex, crosswind, noise, turbulence, and the dither itself. To prevent this anomaly, we consider the constrained optimization program
\begin{equation}\label{eq:constrained_optimization_problem}
p^* =  \arg\max_{p} \; C_l(p) \text{ such that } h(p) \geq 0\,,
\end{equation}
where the unknown function $h(p)$ is to be determined. This constraint must be enforced not only at steady state but also during the transient response. 

\subsection{Safe Flight}
If the constraint of (\ref{eq:constrained_optimization_problem}) holds, $p$ remains within the safe set $S\coloneqq\{p : h(p) \geq 0\}$. Noise, disturbances, and unmodeled dynamics in $h$ might drive $p$ onto the unsafe set. When this situation arises, the priority is to fly the aircraft back to safety. To this end, we adopt the safe ES design framework of~\cite{alan_semiglob_safeESC_2025,alan_local_assign_safeESC}, which requires the following assumption on the constraint map:
\begin{assume}\label{assume:constraint}
    (i) the map $h : \R^2 \to \R$ is continuously differentiable and (ii) $\nabla h(p) \neq 0$ on the set $\{p : h(p) \leq 0\}$.
\end{assume}

Notably, the framework does not require the maximizer itself to be safe. If $p^*$ lies outside $S$, the parameter converges to a neighborhood of the constrained maximizer near the boundary $\partial S \coloneqq \{p : h(p) = 0\}$, oscillating about $\partial S$ due to the dither, with a violation of safety that shrinks with the design constants~\cite{alan_local_assign_safeESC}. If the maps are stationary or the limit cycle persists for too long, the analyst might suppress it using the amplitude modulation logic presented in Section~\ref{sec:amp_mod}. When wind gusts or noise drive the aircraft into the unsafe domain, the framework guarantees a return to $S$ at an assignable rate. However, (ii) cannot be verified in practice since $h$ is unknown. Furthermore,
the safety correction vanishes at the unsafe locations where $\nabla h(p) = 0$, thereby precluding a safe recovery.

The CBF-QP safety filter in~\cite{ames2016_CBFQP} developed for the safety constraint $\dot{h}+ch\geq 0$ with $c>0$, admits the closed-form solution 
\begin{equation}
u =u_0+\frac{\nabla h}{\|\nabla h\|^2}\max\{0,-\nabla h^\top u_0-ch\},
\end{equation}
where $u$ is the update law that deviates the least from the nominal rule $u_0$. This equation, implemented as in~\cite{alan_local_assign_safeESC} with~\eqref{eq:hdef}, leads to 
\begin{align}\label{eq:safe_update}
    \dot{\hat p} = u_0
    + \Gamma_h\,\mathrm{max}_{\delta}\left\{-u_0^\top\hat G_h
    - c\eta_h\right\}\hat{G}_h\,,
\end{align}
where $u_0$ is the unconstrained update law given in (\ref{eq:clampdyn_update}), $\eta_h$ is given as
\begin{align}
    \dot\eta_h = \omega^h_h\, (h-\eta_h)\,,
\end{align} 
with $\omega^h_h > 0$, and the gradient estimate $\hat G_h$ of the constraint $h$ is given by
\begin{align}\label{eq:hat_G_h_dot}
    \dot{\hat G}_h = \omega_l^h\left(\tilde h M(t) - \hat G_h\right)\,,
\end{align}
where $\tilde h = h - \eta_h$ and $\omega_l^h > 0$. $\Gamma_h \in \R$ is given from the Riccati filter
\begin{align}\label{eq:Gamma_h}
    \dot \Gamma_h = \omega_r^h \Gamma_h \left(1 - \Gamma_h\|\hat G_h\|^2\right)\,,
\end{align}
where $\omega_r^h > 0$, which, as in~\eqref{eq:riccati_newton}, has an unstable equilibrium point at $\Gamma_h= 0$ and a stable point at $\Gamma_h = \|\hat G_h\|^{-2}$. Finally, $\mathrm{max}_\delta\{\cdot\}$ is the soft max function\footnote{$\mathrm{max}_\delta\{s\} \coloneqq \tfrac{1}{2}\bigl(s + \sqrt{s^2+\delta}\bigr)$ where ${\delta > 0}$.}. Note that $\omega^h_h$ must be chosen to filter unwanted noise while maintaining a low enough lag to guarantee safety.

In this paper, we use the vortex-induced rolling moment to define both the objective and the constraint. In particular, we consider
\begin{align}\label{eq:hdef}
    h(p) \coloneqq C_l(p) - \underline{C}_l\,,
\end{align}
where $\underline{C}_l > 0$ is the buffer parameter. The chosen value of $\underline{C}_l$ should \jack{be large enough to cover the safety violations} that might be caused by the dither. The extent of such violations depends on the amplitude of the dither, the filters' parameters, and the map $h(p)$. While hard safety guarantees are not possible due to the unknown map, practical safety~\cite{alan_local_assign_safeESC} can be pursued by setting suitable values for $\underline{C}_l$ according to simulation and flight experiments, e.g., if $h(p)= C_l(p)$ is the hard safety constraint, use (\ref{eq:hdef}) with a sufficiently large buffer value. This soft notion of safety is referred to as practical safety.


Our constraint choice makes the unconstrained ES algorithm~\eqref{eq:clampdyn_update} naturally avoid the unsafe region. Crosswind, turbulence, or a maneuvering leading aircraft can driving $p$ into the unsafe domain. Because the baseline update law contains no explicit safety constraints, the aircraft might not be able to exit the unsafe region sufficiently fast. By embedding a safety constraint into the update law we seek to speed up the exiting process. However, as expected, the ability to track a moving vortex and to keep the flight safe is limited by the control authority of the aircraft.



This law enforces practical safety by driving the average value $C_l$ towards the safe domain at a rate $c$. Our constraint choice makes $S$ a level set of the objective map. This property makes $p^*$ an interior point of the safe set, thereby making the second term of the right hand side of (\ref{eq:safe_update}) inactive at the optimum.

\begin{algorithm}[t]
\caption{Threshold-Gated Amplitude Modulation}
\label{alg:amp_mod}
\begin{algorithmic}[1]
\REQUIRE threshold $\overline{C}_l$, dwell $T$, ramp times $T_{\mathrm{down}}, T_{\mathrm{up}}$, sample period $\Delta t$
\STATE $\sigma \leftarrow 1$, $\tau_a \leftarrow 0$
\WHILE{running}
    \IF{$\eta_\sigma(t) \ge \overline{C}_l$}
        \STATE $\tau_a \leftarrow \tau_a + \Delta t$
    \ELSE
        \STATE $\tau_a \leftarrow 0$
    \ENDIF
    \IF{$\tau_a \ge T$}
        \STATE $\sigma \leftarrow \max(0,\ \sigma - \Delta t/T_{\mathrm{down}})$
    \ELSE
        \STATE $\sigma \leftarrow \min(1,\ \sigma + \Delta t/T_{\mathrm{up}})$ 
    \ENDIF
\ENDWHILE
\end{algorithmic}
\end{algorithm}





\subsection{Amplitude Modulation and Plateau Seeking}\label{sec:amp_mod}

The dither required to learn the gradient/Hessian of the unknown functions not only degrades flying qualities but also increases fuel consumption. Furthermore, the response after ES converges exhibits a residual oscillation about $p^*$~\cite{ghaffari_newton_MV_2012}. This oscillation increases fuel consumption without rendering a drag reduction, thereby becoming counter-productive. As such, we modulate and eventually eliminate the excitation as a sufficiently large objective value is reached, i.e., as $p$ approaches the plateau $\{p : C_l(p) \ge \overline{C}_l\}$ when $\overline{C}_l>\underline{C}_l$. Once the plateau is reached, the learning algorithm becomes dormant, only restarting once a moving vortex yields $C_l(p) < \overline{C}_l$ at the converged $p$.  The chosen value for $\overline{C}_l$ should trade off the reduction in drag resulting from flying close to the maximum $p^*$ and the fuel consumption required to persistently excite the system.

This logic is governed by a gain $\sigma \in [0,1]$ computed in Algorithm~\ref{alg:amp_mod}. The value of $\sigma(t)$ decreases linearly in time once $p$ reaches the plateau, and it increases linearly in time once $p$ leaves the plateau. Algorithm 1 adjusts the amplitude of the dither according to a low-pass value of $C_l$, $\eta_\sigma$, as follows: 
\begin{align}\label{eq:amp_mod_filter}
    \dot{\eta}_\sigma = \omega_l^\sigma\,(C_l-\eta_\sigma)\,,
\end{align}
where $\omega_l^\sigma > 0$. As such, measurement noise and the dither-induced oscillation cannot chatter the dwell timer. $\sigma$ scales the excitation and modulates the update according to
\begin{align}
    p_\mathrm{cmd}(t) &= \hat{p}_{\text{mod}}(t) + \sigma(t)\,S(t)\label{eq:mod_update},\\
    \dot{\hat p}_{\text{mod}} &= \andrew{\sigma(t)\,\dot{\hat{p}}}\,,
\end{align}
where $\dot{\hat{p}}$ is the right-hand side of \eqref{eq:safe_update}. Furthermore, it modifies the estimators in \eqref{eq:LP_classic}, \eqref{eq:hess}, and \eqref{eq:Gamma_h} to
\begin{align}
    \dot{\hat G} &= \omega_l\left(\sigma(t)\tilde{C}_l M(t) - \hat G\right)\label{eq:modulated_grad},\\
    \dot{\hat{H}} &= \sigma(t)\,\omega_l^H\left(\tilde{C}_l N(t) - \hat H\right)\label{eq:modulated_newt},\\
    \dot\Gamma_h &= \sigma(t)\,\omega_r^{h}\,\Gamma_h\left(1-\Gamma_h\|\hat G_h\|^2\right)\,,
\end{align}
with~\eqref{eq:hat_G_h_dot} modified in the same manner as~\eqref{eq:modulated_grad}.

When $p$ stays in the plateau for the prescribed dwell time and $\sigma(t)=0$, both the commanded dither and the learning stop. At this state $\hat H$ and $\Gamma_h$ hold their last value whereas $\hat G \rightarrow 0$. When a moving vortex makes $C_l$ drop below $\overline{C}_l$, $p$ leaves the plateau and both the dither and the learning resume. 

Hence, the chosen value for $\overline{C}_l$ not only determines the set of $p$ values to which the optimization eventually converges, but also prescribes the limit after which $p^*$ is no longer pursued. As such, the value of $\overline{C}_l$ determines the trade off posed by wanting to save fuel by reaching a reduced drag configuration against wanting to save fuel by eliminating dithering/actuation. The dither experienced by the aircraft is driven to zero by virtue of a zero gain, but the amplitudes $a_y$ and $a_z$ entering the demodulation signals $M(t)$, $N(t)$ and the underlying equations are fixed, thereby avoiding any singularities.

\begin{figure}[t!]
  \centering
  \begin{subfigure}[t]{0.49\textwidth}
    \centering
    \includegraphics[width=.87\linewidth]{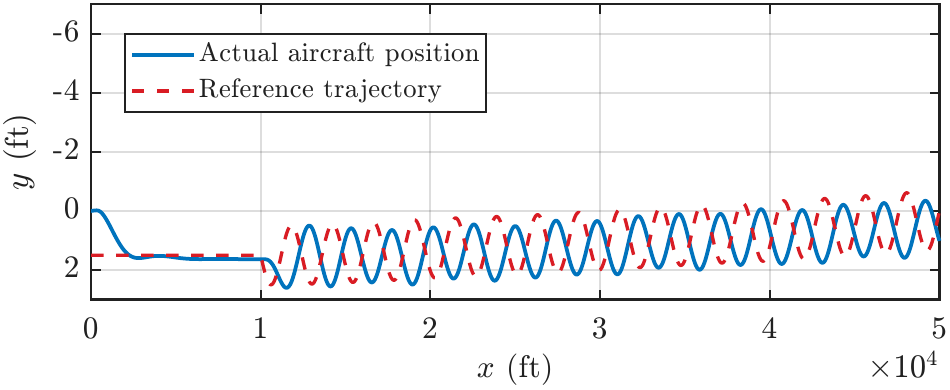}
    \caption{Classical ES with $(k_y\,,k_z) = (0.026,0.027)$.}
    \vspace{0.14cm}
    \label{fig:slow_classic}
  \end{subfigure}\hfill
  \begin{subfigure}[t]{0.49\textwidth}
    \centering
    \includegraphics[width=.87\linewidth]{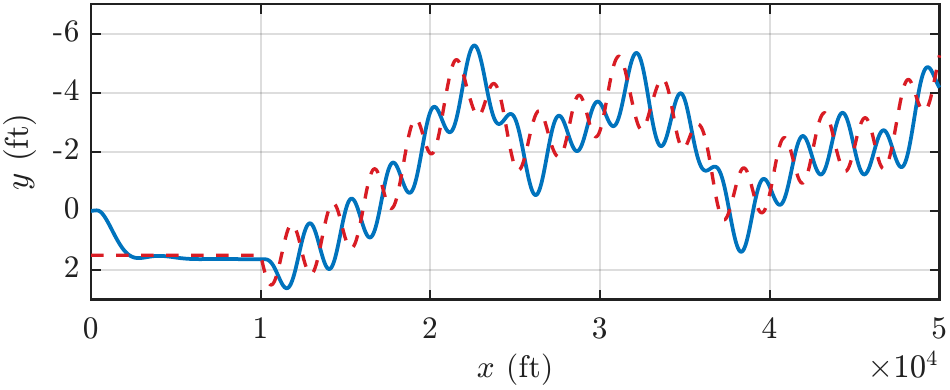}
    \caption{Classical ES with $(k_y\,,k_z) = (0.38,0.38)$.}
    \vspace{0.14cm}
    \label{fig:fast_classic}
  \end{subfigure}\hfill
  \begin{subfigure}[t]{0.49\textwidth}
    \centering
    \includegraphics[width=.87\linewidth]{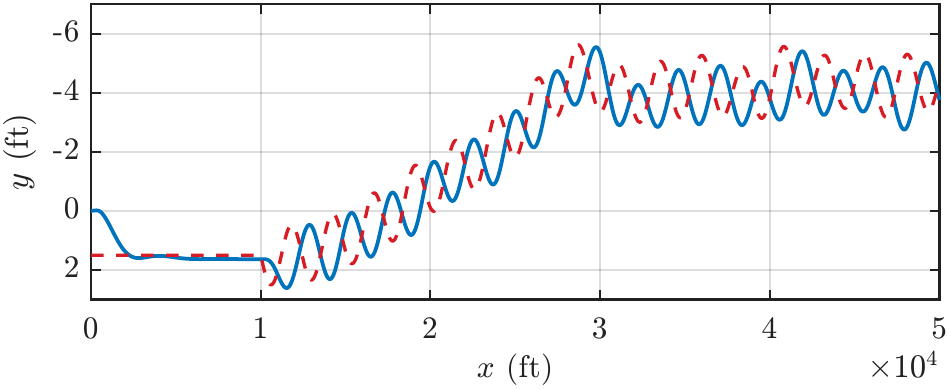}
    \caption{Newton-based ES with $(k_y\,,k_z) = (0.026,0.027)$, $\Gamma(0) =\mathrm{diag}(-7,\,-7.5)$, and $\bar{\gamma} = 10$.}
    \label{fig:newton}
  \end{subfigure}
  
  \caption{Comparison between classical and Newton-based ES. Fig.~\ref{fig:slow_classic} and Fig.~\ref{fig:fast_classic} depict classical ES with slow and fast learning gains, respectively. The former exhibits impractically slow convergence, while the latter suffers from overshoot due to the asymmetric map (see Section~\ref{sec:classical_ES}). In contrast, the Newton-based ES  (Fig.~\ref{fig:newton}) converges rapidly without overshoot.}
  \label{fig:f16-sim_classic_vs_newton}
\vspace{-0.5cm}
\end{figure}

\section{F-16 Formation flying}\label{sec:sim}
In the numerical examples that follow we apply APOGEE to the formation flying of an F-16. To this end we use the high-fidelity nonlinear F-16 model detailed in~\cite{morelli2016aircraft} based on~\cite{stevens2015aircraft}. The simulation is constructed on the wind tunnel aerodynamic database for a 16\% scale model of the F-16 aircraft and a ground test database for engine thrust given in~\cite{nguyen1979simulator}. The simulation is run at a sampling rate of $200\,\mathrm{Hz}$ under significant measurement noise and the \andrew{reference position} is set as \andrew{$(y_{\mathrm{ref}},z_{\mathrm{ref}}) = (0,-25000)\,\mathrm{ft}$} and the initial condition is set to $p_0 = (1.5,3)\,\mathrm{ft}$. Furthermore, we use the aerodynamic vortex model obtained experimentally in~\cite{Hansen02}. This model is used to evaluate the instantaneous value of $C_l(p)$, with no other information about the underlying map used for learning. Additionally, to facilitate parameter choices, the measured output $C_l$ is scaled by a factor of $1000$ since the unscaled magnitude is of order $10^{-3}$.

APOGEE dither parameters are chosen as $a_y = a_z = 1\,\mathrm{ft}$, $\omega_y = 1.3\,\mathrm{rad/s}$, $\omega_z = 0.76,\mathrm{rad/s}$, $\beta_y = -2\pi/3\,\mathrm{rad}$, and $\beta_z = -\pi/5\,\mathrm{rad}$. The update gain is $K = \mathrm{diag}(0.026,\,0.027)$, and the Hessian estimation uses $\omega_r = 0.03$, $\bar\gamma = 10$, and $\Gamma(0) =\mathrm{diag}(-7,\,-7.5)$.
The safety filter parameters use $\underline{C}_l = 3.3$, $\omega_r^h = 0.15$, $c = 2$, $\Gamma_h(0) = 1$, and $\delta = 0.01$, and the amplitude modulation uses $\bar{C}_l = 9.2$, $T = 3\,\mathrm{s}$, and $T_{\mathrm{down}} = 6\,\mathrm{s},  T_{\mathrm{up}} = 4.5\,\mathrm{s}$. The filter frequencies are $\omega_h  = \omega_l^\sigma = \omega_l = \omega_l^{H} = \omega_l^h = 0.5$ and $\omega^h_h = 6$ all in $\mathrm{rad/s}$. APOGEE is turned on at $20\,\mathrm{s}$, time when the response reaches a steady state.

\paragraph{Accelerating convergence}
Fig.~\ref{fig:f16-sim_classic_vs_newton} compares the classic ES and the Newton-based ES on a stationary vortex. The amplitude modulation was not simulated in this case and only ground tracks are shown, as the map asymmetry is most severe in the lateral direction and the overshoot is clearly displayed. Classical ES forces a choice between two poor gain options: one tuned for the shallow side in which the map crawls, and one tuned more than tenfold (comparable to $\bar\gamma$) faster, in which the aircraft overshoots from the steep side. The Newton update with the clamped inverse-Hessian estimate converges rapidly without overshoot and adapts to the unknown curvature.

\begin{figure}[t!]
  \centering
  \begin{subfigure}[t]{0.49\textwidth}
    \centering
    \includegraphics[width=.87\linewidth]{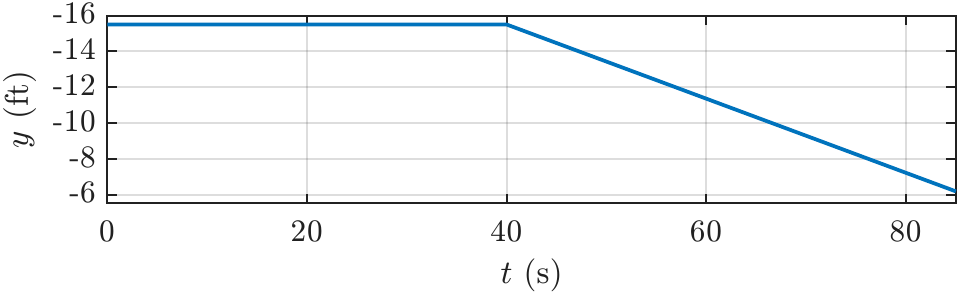}
    \caption{Vortex lateral position over time.}
    \vspace{0.15cm}
    \label{fig:vortex_pos_cbf}
  \end{subfigure}\hfill
  \begin{subfigure}[t]{0.49\textwidth}
    \centering
    \includegraphics[width=.87\linewidth]{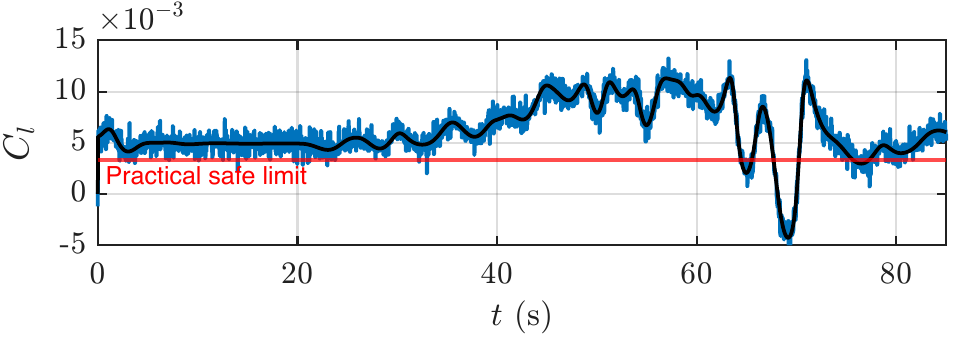}
    \caption{Newton-based ES without safety filter.}
    \vspace{0.15cm}
    \label{fig:no_filter}
  \end{subfigure}\hfill
  \begin{subfigure}[t]{0.49\textwidth}
    \centering
    \includegraphics[width=.87\linewidth]{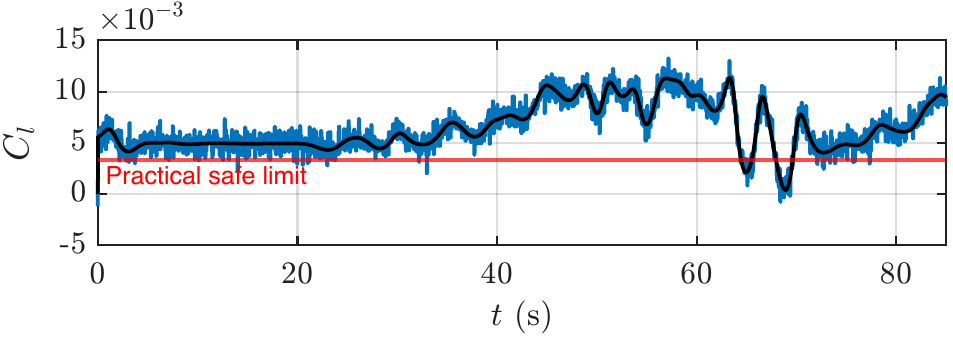}
    \caption{Newton-based ES with safety filter.}
    \label{fig:filter_on}
  \end{subfigure}

  \caption{Newton-based ES with and without the safety filter when the vortex moves towards the aircraft (Fig.~\ref{fig:vortex_pos_cbf}). Blue, black, and red lines in the $C_l$ time histories denote the noisy measurement of $C_l$, the true $C_l$, and the practical safety limit $\underline{C}_l = 3.3\times 10^{-3}$ respectively. 
  }
  \label{fig:f16-sim_filter_vs_no_filter}
\vspace{-0.4cm}
\end{figure}

\paragraph{Avoiding unsafe flight}
Fig.~\ref{fig:f16-sim_filter_vs_no_filter} shows the responses when the vortex moves laterally towards the aircraft, resulting in an unsafe state. The unconstrained update, shown in Fig.~\ref{fig:no_filter}, results in $C_l$ falling below the hard safety limit of zero. In contrast, the constrained update triggered by the practical safety limit $C_l=\underline{C}_l$, shown in Fig.~\ref{fig:filter_on}, prevents loss of control by ensuring $C_l>0$.

\begin{figure*}[t!]
  \centering
  \includegraphics[width=.88\linewidth]{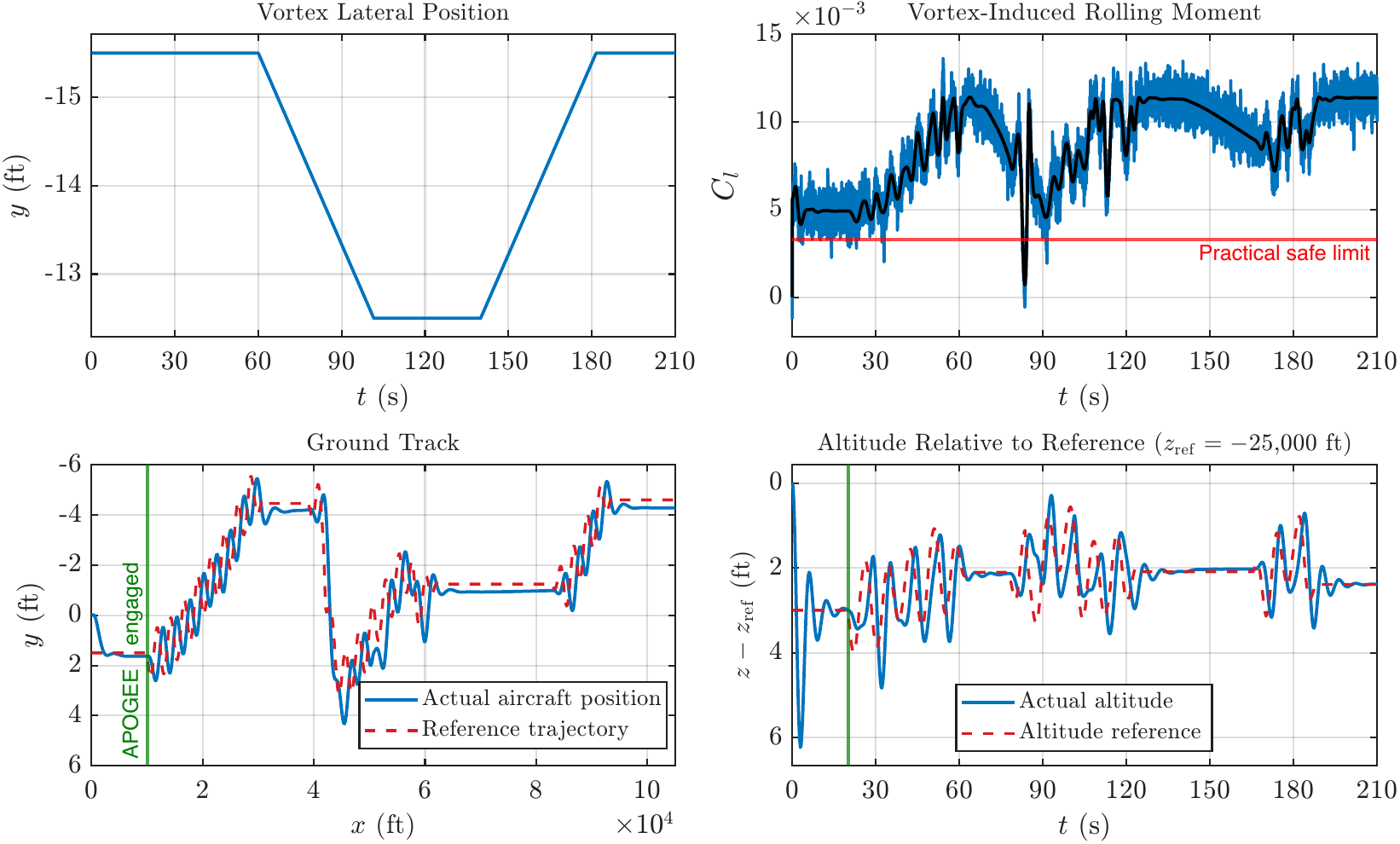}
    \caption{High-fidelity F-16 simulation starting from $p_0 = (1.5,3)\,\mathrm{ft}$. Top left: lateral position of a moving vortex. Top right: true and measured vortex-induced rolling moment. Bottom left: relative lateral command and response for \andrew{$y_\mathrm{ref} = 0\,\mathrm{ft}$}. Bottom right: relative altitude command and response for \andrew{$z_\mathrm{ref} = -25,000\,\mathrm{ft}$}. APOGEE is turned on at $t = 20\,\mathrm{s}$.}
  \label{fig:f16-sim}
\vspace{-0.5cm}
\end{figure*}

\paragraph{Full operation}
Fig.~\ref{fig:f16-sim} shows an APOGEE trajectory with a translating vortex and all components active. Once $p$ reaches the first plateau, Algorithm~\ref{alg:amp_mod} ramps the excitation to zero, and the aircraft holds position. As the vortex translates laterally, the resulting reduction of $C_l$ resumes the seeking process, alternating between seeking and dither-free station keeping. When the vortex moves toward the aircraft, $C_l$ drops sharply and the safety filter drives $p$ outside the unsafe set, enforcing the practical safe limit.

\section{Conclusion}
This paper presents APOGEE, a model-free guidance architecture for drag reduction in formation flight, built from ES on the measured vortex-induced rolling moment. The Newton-based update is modified by a spectral clamp on the inverse-Hessian estimate, bounding the transient gain on any asymmetric wake map. A practical CBF-QP safety filter acting on the same measurement keeps the response away from the \andrew{unsafe} region, while an amplitude modulation algorithm turns the learning on and off depending on the system reaching a point of acceptable performance. High-fidelity F-16 simulations are used to demonstrate the efficacy and safety of APOGEE. In future work, we will consider other safety constraints, evaluate the domain where the algorithm remains both effective and safe, and develop a means to prescribe robust values to the tunable parameters.

\vspace{-0.2cm}
\bibliographystyle{IEEEtranS}
\bibliography{refs}

\end{document}